# Interatomic forces, phonons, the Foreman-Lomer Theorem and the Blackman Sum Rule

A. M. STEWART

*Emeritus Faculty*
*The Australian National University,*
*Canberra, ACT 0200, Australia.*

*Abstract*

Foreman and Lomer proposed in 1957 a method of estimating the harmonic forces between parallel planes of atoms of primitive cubic crystals by Fourier transforming the squared frequencies of phonons propagating along principal directions. A generalized form of this theorem is derived in this paper and it is shown that it is more appropriate to apply the method to certain combinations of the phonon dispersion relations rather than to individual dispersion relations themselves. Further, it is also shown how the method may be extended to the non-primitive hexagonal close packed and diamond lattices. Explicit, exact and general relations in terms of atomic force constants are found for deviations from the Blackman sum rule which itself is shown to be derived from the generalized Foreman-Lomer theorem.

## 1. Introduction
The theory of the lattice dynamics of perfect crystals, see Maradudin *et al.*[1] and Srivastava[2] for reviews and standard notation, has traditionally been discussed in terms of atomic force constants, the force constants between different atoms in the lattice. The model is exact, in the sense that given a complete knowledge of the eigenvalues and eigenfunctions of the phonon spectrum a complete and unique set of force constants may be determined, for example from equation (5) of this paper. From this atomic force model an extensive understanding of the symmetry properties of phonons has been developed[1,3,4]. However it is far from clear that these formally constructed force constants are the best way of representing all the interactions in the crystal lattice that give rise to its dynamics, particularly in metals. Modern approaches to lattice dynamics[5,6] rely on *ab initio* methods such as density functional theory, the frozen lattice approximation etc. to calculate phonon energies with excellent agreement with experiment.

However, it is also far from clear that the problems of self-assembly and dynamics of more complex materials in the amorphous, colloidal and biological realms can be approached in the same way. It seems possible that the force constant model may be of use as a first step in investigating the properties of these increasingly important materials. Accordingly it seems of value to explore as far as possible the force constant model of perfect crystals. Perfect crystals are of interest because experimental information about atomic forces is obtained from them most easily. Only when a satisfactory understanding has been obtained of atomic forces in perfect crystals can the theory be applied with confidence to the more complex materials.

As will be seen, all the harmonic force components between atoms in crystals may in principle be obtained if the phonon eigenfrequencies and eigenvectors are known at every point in the Brillouin zone[7]. However the phonon eigenvectors, the magnitude and direction of vibrational motion of every atom in the unit cell, are not easy to measure[8] and experimental results are most often only made available in the form of a vibrational spectrum, a plot of the several branches of the resonant frequencies of the lattices at various excitation wavelengths.

Because of this limitation the procedure that has often been adopted to analyze data is to assume that only a restricted set of force constants have non-zero values (those connecting the neighbours in the vicinity of the atom in question) and adjust these to get the best fit to the vibrational spectrum. The question then arises as to how many neighbours need to be taken account of to fit this spectrum. While this may be determined by trial and error, a more systematic procedure is desirable. Foreman and Lomer[7] (see Kunc and Martin[9] for a different approach) have provided a partial answer to this need by arguing that, for the phonons propagating along the symmetry axes of monatomic cubic crystals, the Fourier series expansions of the squares of the eigenfrequencies give directly certain force constants between entire planes of atoms, each term in the Fourier series being determined by the interactions between corresponding planes. This approach has been applied to the metals aluminium[4,7,10], niobium[11], lead[12] and sodium[13]. In some cases, the interactions are found to be of very long range, up to fifteenth nearest neighbours needing to be taken account of.

In this paper a general derivation of the Foreman-Lomer theorem is demonstrated[14,15] and it is shown how more detailed information about the range of atomic forces in crystals may be obtained from simple combinations of individual branches of the dispersion relations for certain directions not only of primitive cubic crystals but also of the non-primitive hexagonal close packed and diamond lattices. It is also shown how a more general statement of the Blackman sum rule[16] arises naturally out of the theory. The general version of this rule states that the range of non-Coulombic atomic forces in a crystal of *any* structure can be deduced from the wave-vector dependence of the sum of the squares of the phonon frequencies. In section **2** of this paper the rudimentary ideas of lattice dynamics needed for this work are summarized. In section **3** the key assumption of the work is stated, namely how the sum over lattice points in the dynamical matrix may be carried out by first summing over individual planes of atoms perpendicular to a particular reciprocal lattice vector and second summing over all these planes. This leads to the generalization of the Foreman-Lomer theorem. In section **4** it is shown how the generalized Blackman sum rules arises out of the same theory; previously it had not been appreciated that there was a connection between these two ideas. In sections **5** and **6** it is demonstrated how these arguments may be applied to primitive and non-primitive crystal lattices respectively. Finally, although the bulk of the paper is a review of some exact results arising from simple models of perfectly periodic solids, some comments are made in section **7** about the applicability of the methods used to the dynamical modes of non-periodic solids and their thermodynamic properties. A short summary **8** ends the paper.

## 2. Rudiments of the Lattice Dynamics of Perfect Crystals

We consider a perfectly periodic lattice whose lattice vectors are given by

$$\boldsymbol{T}(l_1,l_2,l_3) = l_1\boldsymbol{a}_1 + l_2\boldsymbol{a}_2 + l_3\boldsymbol{a}_3 \quad , \qquad (1)$$

where the $\boldsymbol{a}_i$ are the basis vectors of the direct lattice and the $l_i$ are integers. In each cell there is a basis of $r$ atoms situated at their equilibrium positions denoted by $\kappa$ with $1 \le \kappa \le r$. A particular atom in the crystal is identified by $(l,\kappa)$ where $l$ is a member of the set of $l_i$ that specifies the cell and $\kappa$ describes the atom within that cell. If the atom at $(l',\kappa')$ is displaced from its equilibrium position by $\delta x^\beta(l',\kappa')$ in the $\beta$ direction where $\beta$ is one of the Cartesian directions *x, y, z* then in the harmonic approximation the equations for the collective motion of the atoms at frequency $\omega$ are given in terms of the (Fourier transformed) dynamical matrix **D** by[1,2]

$$\sum_{\kappa'=1}^{r} \sum_{\beta=1}^{3} \{D_{\kappa\kappa'}^{\alpha\beta}(\boldsymbol{q}) - \omega^2(\boldsymbol{q})\delta(\alpha,\beta)\delta(\kappa,\kappa')\}u^\beta(\kappa') = 0 \quad , \qquad (2)$$

where

$$D_{\kappa\kappa'}^{\alpha\beta}(\boldsymbol{q}) = (M_\kappa M_{\kappa'})^{-1/2} \sum_{l'} \Phi^{\alpha\beta}(l,\kappa;l',\kappa')\exp[-i\boldsymbol{q}\cdot\{\boldsymbol{T}(l) - \boldsymbol{T}(l')\}] \quad , \qquad (3)$$

the $\Phi^{\alpha\beta}(l,\kappa;l',\kappa')$ being the harmonic force tensor and $\delta(\alpha,\beta)$ etc. the Kronecker delta. The atomic displacement $\delta x^\alpha$ in the direction $\alpha$ as a function of time *t* is

$$\delta x^\alpha(l,\kappa) = \exp[i\{\boldsymbol{q}\cdot\boldsymbol{T}(l_1,l_2,l_3) - \omega(\boldsymbol{q})t\}]u^\alpha(\kappa)/\sqrt{M_\kappa} \quad . \qquad (4)$$

The dynamical matrix is periodic in *q* and need be defined only in the first Brillouin zone[17]. The dynamical matrix is a tensor under rotation in the indices $\alpha, \beta$ and has matrix indices $\kappa, \kappa'$. The inverse relation is:

$$\Phi^{\alpha\beta}(l,\kappa;l',\kappa') = \frac{(M_\kappa M_{\kappa'})^{1/2}}{N_c} \sum_{\boldsymbol{q}}^{BZ} D_{\kappa\kappa'}^{\alpha\beta}(\boldsymbol{q})\exp[i\boldsymbol{q}\cdot\{\boldsymbol{T}(l) - \boldsymbol{T}(l')\}] \quad , \qquad (5)$$

where $N_c$ is the number of cells in the crystal of volume *V* and the sum over wave-vectors is restricted to the first Brillouin zone.

Each of the 3*r* eigenvalues of equation (2) at a wave-vector *q*, denoted by $\omega_i^2(\boldsymbol{q})$, with index *i*, $1 \leq i \leq 3r$, form a continuous branch as a function of *q*. The 3*r* eigenvectors $|i(\boldsymbol{q})\rangle$ may be expressed formally as:

$$|i(\boldsymbol{q})\rangle = \sum_{\kappa=1}^{r} \sum_{\alpha=1}^{3} |\kappa\alpha\rangle\langle\kappa\alpha|i\rangle \quad , \qquad (6)$$

where the $|\kappa\alpha\rangle$ represent the $u^\alpha(\kappa)$ and the $\langle\kappa\alpha|i\rangle$ are the expansion coefficients of the eigenvectors. Dirac's bracket notation is used for convenience. In general the coefficients $\langle\kappa\alpha|i\rangle$ will be functions of *q* and will depend on the numerical values of the force constants, but in some special cases that will interest us in this paper their dependence on *q*, if any, is determined entirely by requirements of symmetry.

The dynamical matrix may diagonalised by the transformation (6) to give:

$$\omega_i^2(\boldsymbol{q}) = \sum_{\kappa\alpha} \sum_{\kappa'\beta} \langle\kappa\alpha|i\rangle\langle i|\kappa'\beta\rangle D_{\kappa\kappa'}^{\alpha\beta}(\boldsymbol{q}) \quad , \qquad (7)$$

which may be written in matrix form $\boldsymbol{E} = \boldsymbol{U}\boldsymbol{D}\boldsymbol{U}^{-1}$ where $\boldsymbol{E}$ is the diagonal matrix of the eigenvalues $\omega_i(\boldsymbol{q})^2$ and $\boldsymbol{U}$ is the eigenfunction matrix of the coefficients $\langle i|\kappa\alpha\rangle$. The inverse transformation is:

$$D_{\kappa\kappa'}^{\alpha\beta}(\boldsymbol{q}) = \sum_i \langle \kappa'\beta|i\rangle\langle i|\kappa\alpha\rangle \omega_i^2(\boldsymbol{q}) \qquad , \qquad (8)$$

or in matrix form $\boldsymbol{D} = \boldsymbol{U}^{-1}\boldsymbol{E}\boldsymbol{U}$. All the harmonic force constants may then be obtained first by getting $\boldsymbol{D}$ from (8) and then $\Phi$ from (5) provided that the $\langle \kappa\alpha|i\rangle$ are already known.

From the definition of the inner product the following relation holds: $\langle i|\kappa\alpha\rangle = \langle \kappa\alpha|i\rangle^*$, and from its definition in equation (3) and the symmetry of the force tensor $\Phi^{\alpha\beta}(l,\kappa;l',\kappa') = \Phi^{\beta\alpha}(l',\kappa';l,\kappa)$ the dynamical matrix has the properties: $\text{Re}\{D_{\kappa\kappa'}^{\alpha\beta}(\boldsymbol{q})\} = \text{Re}\{D_{\kappa'\kappa}^{\beta\alpha}(\boldsymbol{q})\}$ and $\text{Im}\{D_{\kappa\kappa'}^{\alpha\beta}(\boldsymbol{q})\} = -\text{Im}\{D_{\kappa'\kappa}^{\beta\alpha}(\boldsymbol{q})\}$. Accordingly $\boldsymbol{D}$ is a Hermitian matrix and has real eigenvalues. If any of these are negative this indicates that the lattice is unstable and a spontaneous transition is expected to a lattice of different structure. It follows from (7) that because of the existence of the sum over both $\alpha\kappa$ and $\beta\kappa'$ that:

$$\omega_i^2(\boldsymbol{q}) = \sum_{\kappa\alpha}\sum_{\kappa'\beta}[\text{Re}\{\langle\kappa\alpha|i\rangle\langle i|\kappa'\beta\rangle\}\text{Re}\{D_{\kappa\kappa'}^{\alpha\beta}(\boldsymbol{q})\} - \text{Im}\{\langle\kappa\alpha|i\rangle\langle i|\kappa'\beta\rangle\}\text{Im}\{D_{\kappa\kappa'}^{\alpha\beta}(\boldsymbol{q})\}], \quad (9)$$

a real quantity as expected. There is another symmetry property of the force tensor that will be needed. If the every atom in the crystal is translated by the same displacement then the net force on any atom must remain zero. Hence[1]

$$\sum_{l,\kappa} \Phi^{\alpha\beta}(l,\kappa;l',\kappa') = 0 \qquad . \qquad (10)$$

**3. Summing the Dynamical Matrix over Planes of Atoms to get the Foreman-Lomer Theorem.**

When carrying out the sum over $l'$ to evaluate the dynamical matrix $\boldsymbol{D}$ in equation (3) we take the wave-vector $\boldsymbol{q}$ to be proportional to a reciprocal lattice vector $\boldsymbol{G}$

$$\boldsymbol{q} = \eta\boldsymbol{G} \qquad , \qquad (11)$$

where $\eta$ is a number and

$$\boldsymbol{G}(h_1,h_2,h_3) = h_1\boldsymbol{b}_1 + h_2\boldsymbol{b}_2 + h_3\boldsymbol{b}_3 \qquad . \qquad (12)$$

The $h_i$ are integers and the $\boldsymbol{b}$ are the basis vectors of the reciprocal lattice given by

$$\boldsymbol{a}_l \cdot \boldsymbol{b}_m = 2\pi\delta(l,m) \qquad . \qquad (13)$$

$\boldsymbol{a}_l$ being the basis vectors of the direct lattice. Any $\boldsymbol{q}$ in the first Brillouin zone may be obtained from an appropriate $\boldsymbol{G}$, but usually, as in more conventional treatments, high symmetry directions will be considered which correspond to small values of the $h_i$ in equation (12). For $\eta = 0$, $\boldsymbol{q}$ will be at the zone centre, for $\eta = 1/2$ and a high symmetry $G$, $q$ will be at a zone boundary. The plane in real space defined by $T.G = 2\pi(l_1 h_1 + l_2 h_2 + l_3 h_3) = 2\pi N, -\infty < N < \infty$ is perpendicular to $G$. For $N = 0$ it passes through the origin ($T = 0$), for $N = 1$ it is the next crystallographically equivalent plane in the direction of $G$, for $N = -1$ it is the next one in the direction of $-G$ and so forth. The distance between these planes in reciprocal space is[17] $2\pi/|G|$.

The sum in equation (3) over $T(l_1,l_2,l_3)$ may be partitioned into a sum over $(l_1,l_2,l_3)$, subject to the condition $l_1 h_1 + l_2 h_2 + l_3 h_3 = N$ followed by a sum over all the integers $N$, so that[14,15]:

$$D^{\alpha\beta}_{\kappa\kappa'}(\mathbf{q}) = \sum_{N=-\infty}^{+\infty} \phi^{\alpha\beta}_{\kappa\kappa'}(N) \exp(i 2\pi\eta N) \quad , \quad (14)$$

where

$$\phi^{\alpha\beta}_{\kappa\kappa'}(N) = (M_\kappa M_{\kappa'})^{-1/2} \sum_{l_1,l_2,l_3} \Phi^{\alpha\beta}(0,\kappa;l_1,l_2,l_3,\kappa') \delta(N, l_1 h_1 + l_2 h_2 + l_3 h_3) \quad . \quad (15)$$

The $\phi$ coefficients have the symmetry relations $\phi^{\alpha\beta}_{\kappa\kappa'}(-N) = \phi^{\beta\alpha}_{\kappa'\kappa}(N)$ which is a consequence of the relation $\Phi^{\alpha\beta}(\lambda,\kappa;\lambda',\kappa') = \Phi^{\beta\alpha}(\lambda',\kappa';\lambda,\kappa)$, together with translational invariance. It also follows from (10) that

$$\sum_{\kappa'} \sum_{N=-\infty}^{+\infty} \phi^{\alpha\beta}_{\kappa\kappa'}(N) = 0 \quad . \quad (16)$$

Using de Moivre' theorem we get

$$D^{\alpha\beta}_{\kappa\kappa'}(\mathbf{q}) = \phi^{\alpha\beta}_{\kappa\kappa'}(0) + \sum_{N=1}^{+\infty} \{[\phi^{\alpha\beta}_{\kappa\kappa'}(N) + \phi^{\beta\alpha}_{\kappa'\kappa}(N)]\cos(2\pi\eta N) + i[\phi^{\alpha\beta}_{\kappa\kappa'}(N) - \phi^{\beta\alpha}_{\kappa'\kappa}(N)]\sin(2\pi\eta N)\} \quad (17)$$

or equivalently, using the identity $\cos 2x = 1 - \sin^2 x$ and equation (14) with $\eta = 0$,

$$D^{\alpha\beta}_{\kappa\kappa'}(\mathbf{q}) = D^{\alpha\beta}_{\kappa\kappa'}(\mathbf{0}) - 2\sum_{N=1}^{\infty} \{\phi^{\alpha\beta}_{\kappa\kappa'}(N) + \phi^{\beta\alpha}_{\kappa'\kappa}(N)\} \sin^2(\pi\eta N)$$
$$+ i\sum_{N=1}^{\infty} \{\phi^{\alpha\beta}_{\kappa\kappa'}(N) - \phi^{\beta\alpha}_{\kappa'\kappa}(N)\} \sin(2\pi\eta N) \quad . \quad (18)$$

Combining equations (9) and (17) we obtain

$$\omega_i^2(\mathbf{q}) = \sum_{\kappa\alpha} \sum_{\kappa'\beta} \mathrm{Re}(<\kappa\alpha|i><i|\kappa'\beta>)[\phi^{\alpha\beta}_{\kappa\kappa'}(0) + \sum_{N=1}^{\infty} \{\phi^{\alpha\beta}_{\kappa\kappa'}(N) + \phi^{\beta\alpha}_{\kappa'\kappa}(N)\}\cos(2\pi\eta N)]$$
$$- \mathrm{Im}(<\kappa\alpha|i><i|\kappa'\beta>)\sum_{N=1}^{\infty} \{\phi^{\alpha\beta}_{\kappa\kappa'}(N) - \phi^{\beta\alpha}_{\kappa'\kappa}(N)\}\sin(2\pi\eta N) \quad . \quad (19)$$

In those cases, to be discussed later, where the $<\kappa\alpha|i>$ are independent of $\mathbf{q}$, this is the exact statement of the Foreman-Lomer theorem. It says that the Fourier series expansions in wave-vector of the squares of the eigenfrequencies give the force constants between entire planes of atoms, denoted by $N$, perpendicular to the propagation direction of the phonon, the Fourier coefficients for each $N$ being linear combinations of the $\phi_{\kappa\kappa'}^{\alpha\beta}(N)$. It can be seen that a knowledge of the dynamical matrix along a line in reciprocal space between the origin and a lattice point allows the harmonic forces between atomic planes perpendicular to that line to be obtained, whereas a knowledge of the dynamic matrix over the whole of the volume of the Brillouin zone is needed to determine the forces between individual atoms through (5).

**4. The Blackman Sum Rule**

By summing equation (19) over the phonon branch index $i$ and using closure $\sum_{i=1}^{3r} |i\rangle\langle i| = 1$, the exact extended form of the Blackman sum rule[14] is obtained:

$$\sum_i \omega_i^2(\mathbf{q}) = \sum_i \omega_i^2(\mathbf{0}) - 4\sum_{N=1}^{\infty} \phi(N)\sin^2(\pi\eta N) \quad , \qquad (20)$$

where $\quad \phi(N) = \sum_{\kappa\alpha} \phi_{\kappa\kappa}^{\alpha\alpha}(N)$ .                                                                        (21)

To derive the constant term in (20) we note that from (19) it comes to $\sum_{\alpha\kappa}\sum_{N=-\infty}^{\infty}\phi_{\kappa\kappa}^{\alpha\alpha}(N)$. When the $\phi$ are expressed in terms of the $\Phi$ with equation (15) then it becomes $\sum_{\alpha\kappa}\sum_{l'}\Phi^{\alpha\alpha}(0,\kappa;l',\kappa)/M_\kappa$ because summing (15) over $N$ is equivalent to summing the $\Phi$ over all $l'_i$. Alternatively from (8) we get

$$\sum_i \omega_i^2(\mathbf{q}) = \sum_{\alpha\kappa} D_{\kappa\kappa}^{\alpha\alpha}(\mathbf{q}) \quad , \qquad (22)$$

and particularly

$$\sum_i \omega_i^2(\mathbf{0}) = \sum_{\alpha\kappa} D_{\kappa\kappa}^{\alpha\alpha}(\mathbf{0}) = \sum_{\alpha\kappa}\sum_{l'}\Phi^{\alpha\alpha}(0,\kappa;l',\kappa)/M_\kappa \quad , \qquad (23)$$

which gives the constant term in (20).

Equation (20) is the extended version of the Blackman Sum Rule[14,15]. In its original form[16] this rule stated that if an ionic crystal has only Coulombic forces between next nearest and more distant neighbours then the sum of the squares of the lattice frequencies is independent of wave-vector, the second term on the right-hand side of (20) vanishing in that case. The extension to the Blackman sum rule is the second term of (20) which shows quantitatively how non-Coulombic forces contribute to the wave-vector dependence of the sum of squared frequencies. The extended form has the advantages of giving a quantitative expression for deviations from the original rule and of being valid for *any* crystal. But both forms of the rule possess the serious drawback that taking the sum over branches loses all the information about Coulomb forces because, while they contribute to each individual element of the dynamical matrix, they make no contribution to its trace[7,16]. The reason for this is that in equations (20, 21) the relation involves the trace over $\alpha$ of the force constants. The latter are the second spatial derivatives of the potentials that give the force. The trace therefore produces a $\nabla^2$ acting on the potential and any potential, such as the Coulomb potential, that obeys Laplace's equation away from the origin will therefore contribute nothing.

The simplest system of forces to stabilize an ionic crystal consists of repulsion between nearest neighbours and Coulombic forces between next and further neighbours. This model would obey the simple form of the Blackman Sum Rule[16]. Deviations from Blackman's sum rule have been discussed by several authors[18-20]. The left-hand side of equation (20) has been calculated along the principal directions of several important non-primitive crystals by Rosenstock[19,20]. He found that silicon and germanium obeyed Blackman's sum rule well but that diamond and gallium arsenide showed significant deviations from it which could be attributed to the presence of non-Laplacian forces between next-nearest and further neighbours.

## 5. Primitive Cubic Lattices

We now consider particular lattices in which the vibrational eigenfunctions are determined by symmetry alone and do not depend explicitly on the force constants[3,4]. In these cases the generalized Foreman-Lomer theorem, equation (19), may be applied directly.

When there is only one atom per unit cell $D$ is a real, symmetric 3x3 matrix that is an even function of $q$ because $\Phi^{\alpha\beta} = \Phi^{\beta\alpha}$ and $\Phi^{\alpha\beta}(T) = \Phi^{\alpha\beta}(-T)$. For the three primitive cubic lattices $D^{\alpha\beta}(q)$ is determined by symmetry requirements alone[3,4]. The phonon eigenvectors for the three principal directions are independent of wave-vector, see for example Willis and Pryor[21]. For phonons propagating in the [001] direction $D^{\alpha\beta}(q)$ is diagonal with elements $D^{zz}(q) = \omega_L^2$ for the longitudinal acoustic (LA) mode and $D^{xx}(q) = \omega_T^2$ and $D^{yy}(q) = \omega_T^2$, the two TA (transverse acoustic) modes being degenerate. For the LO phonon the eigenfunction is $|i\rangle = |z\rangle$ and the squared frequencies may be expressed directly by the Foreman-Lomer method as:

$$\omega_L^2(q) = \phi^{zz}(0) + 2 \sum_{N=1}^{\infty} \phi^{zz}(N)\cos(2\pi\eta N) \qquad , \qquad (24)$$

which becomes

$$\omega_L^2(q) = -4 \sum_{N=1}^{\infty} \phi^{zz}(N)\sin^2(\pi\eta N) \qquad , \qquad (25)$$

as the constant term vanishes by (16). There is a similar expression for the TA modes with $\phi^{xx}$ replacing $\phi^{zz}$. Where there is only one term in the expansion then $\omega_i$ becomes proportional to $\sin(\pi\eta)$, recovering the standard result for the linear chain model[17].

For phonons propagating in the [110] direction there is no degeneracy so the eigenvalue matrix $E$ is[4]

$$E_{110} = \begin{bmatrix} \omega_T^2(q) & 0 & 0 \\ 0 & \omega_L^2(q) & 0 \\ 0 & 0 & \omega_z^2(q) \end{bmatrix} \qquad . \qquad (26)$$

The eigenfunction matrix $U^{-1}$ of the coefficients $\langle\kappa\alpha|i\rangle$ is[4]:

$$U_{110}^{-1} = \frac{1}{\sqrt{2}} \begin{bmatrix} 1 & 1 & 0 \\ -1 & 1 & 0 \\ 0 & 0 & \sqrt{2} \end{bmatrix} \quad , \qquad (27)$$

where the rows denote the $\alpha$ and the columns the $i$. The $i = 1$ mode is TA with the atoms vibrating in the [1,-1,0] direction, the $i = 2$ mode is LA and the $i = 3$ mode is TA with the atoms vibrating in the $z$ direction. The dynamical matrix is obtained from $\boldsymbol{D} = \boldsymbol{U}^{-1}\boldsymbol{E}\boldsymbol{U}$ (8) as:

$$D_{110} = \frac{1}{2} \begin{bmatrix} \omega_L^2 + \omega_T^2 & \omega_L^2 - \omega_T^2 & 0 \\ \omega_L^2 - \omega_T^2 & \omega_L^2 + \omega_T^2 & 0 \\ 0 & 0 & 2\omega_z^2 \end{bmatrix} \quad , \qquad (28)$$

with the $\boldsymbol{q}$ dependence suppressed. Each element of (28) can now be expanded in the way that was done in equation (25) and it can be deduced whether the off-diagonal elements of $\boldsymbol{D}$ have a range that is different to that of the diagonal elements. It would appear to be more fundamental to expand the combinations of eigenvalues shown in each element of (28) in a Fourier series rather than the individual eigenvalues, as has usually been done following Foreman and Lomer[7], because then the behaviour of the different elements of the dynamical matrix can be distinguished.

For phonons propagating along the [111] direction the transverse modes are degenerate and the eigenfunction matrix is[4]:

$$U_{111}^{-1} = \begin{bmatrix} \frac{1}{\sqrt{2}} & \frac{1}{\sqrt{3}} & -\frac{1}{\sqrt{6}} \\ -\frac{1}{\sqrt{2}} & \frac{1}{\sqrt{3}} & -\frac{1}{\sqrt{6}} \\ 0 & \frac{1}{\sqrt{3}} & \frac{2}{\sqrt{6}} \end{bmatrix} \quad , \qquad (29)$$

again the $i = 2$ mode is longitudinal. From the transformation $\boldsymbol{D} = \boldsymbol{U}^{-1}\boldsymbol{E}\boldsymbol{U}$ the dynamical matrix is found to be:

$$D_{111} = \frac{1}{3} \begin{bmatrix} P & Q & Q \\ Q & P & Q \\ Q & Q & P \end{bmatrix} \quad , \qquad (30)$$

where $P = \omega_L^2 + 2\omega_T^2$ and $Q = \omega_L^2 - \omega_T^2$. Again, each element may be expanded in the form of equation (25).

## 6. Non-Primitive Lattices
There exist a few cases among non-primitive lattices in which the phonon eigenvectors are determined by symmetry alone and do not depend on the atomic force constants. In these cases the method of analysis proposed in this paper may be used. The hexagonal close packed and diamond lattices are two examples. The requirements placed by symmetry upon the lattice vibrations of the hexagonal close packed lattice have been investigated by Warren[4]. He found that for phonons propagating along the hexagonal axis in the [001] direction the dynamical matrix was of the form:

$$\boldsymbol{D}_{001} = \begin{bmatrix} A & & & & C & \\ & A & & & & C \\ & & B & & & D \\ C^* & & & A & & \\ & C^* & & & A & \\ & & D^* & & & B \end{bmatrix}, \quad (31)$$

with $C^* = C\exp(i2\pi\eta)$ and $D^* = D\exp(i2\pi\eta)$. At the origin ($\eta = 0$) $C = -A$ and $D = -B$, and the off-diagonal elements vanish on the zone face at $\eta = 1/2$ or $\boldsymbol{q} = \boldsymbol{G}_{001}/2$, where $\boldsymbol{G}_{001}$ is the basis vector of the reciprocal lattice in the [001] direction. In the above expression the first three rows or columns refer to the x, y and z directions associated with the first atomic site and the second three to the second. The twenty-four elements left blank are equal to zero. Warren found the eigenfunction matrix to be:

$$\boldsymbol{U}_{001}^{-1} = \frac{1}{\sqrt{2}} \begin{bmatrix} 0 & 0 & 0 & 1 & 1 & 0 \\ 0 & 0 & 1 & 0 & 0 & 1 \\ 1 & 1 & 0 & 0 & 0 & 0 \\ 0 & 0 & 0 & -d & d & 0 \\ 0 & 0 & -d & 0 & 0 & d \\ d & -d & 0 & 0 & 0 & 0 \end{bmatrix}, \quad (32)$$

where $d = \exp(i\pi\eta)$. The $\eta$ used by Warren is larger than the one used throughout this paper by a factor of two. The modes at $\boldsymbol{q} = \boldsymbol{0}$ are, from left to right: LA, LO, TO, TO, TA, TA (O = optical). Warren found the eigenvalues to be:

$$\begin{aligned} \omega_1^2 &= B + Dd & \text{(LA)} \\ \omega_2^2 &= B - Dd & \text{(LO)} \\ \omega_4^2 &= A - Cd & \text{(2 x TO)} \\ \omega_5^2 &= A + Cd & \text{(2 x TA)} \end{aligned} \quad (33)$$

the transverse modes (3 and 4) and (5 and 6) being doubly degenerate.

From the above relations it follows that $\omega_1^2 + \omega_2^2 = 2B$, $\omega_4^2 + \omega_5^2 = 2A$, $\omega_1^2 - \omega_2^2 = 2Dd$ and $\omega_5^2 - \omega_4^2 = 2Cd$. The two diagonal matrix elements may then each be expressed in a Fourier series:

$$\omega_1^2(\boldsymbol{q}) + \omega_2^2(\boldsymbol{q}) = \omega_{LO}^2(\boldsymbol{q}) - 4\sum_{N=1}^{\infty} \{1 - \cos(2\pi\eta N)\}\phi_{11}^{zz}(N) \quad (34)$$

and

$$\omega_4^2(\boldsymbol{q}) + \omega_5^2(\boldsymbol{q}) = \omega_{TO}^2(\boldsymbol{q}) - 4\sum_{N=1}^{\infty} \{1 - \cos(2\pi\eta N)\}\phi_{11}^{xx}(N), \quad (35)$$

where the first terms on the right hand sides are the squared frequencies at $q = 0$ of the LO and TO modes, given respectively for the LO mode by $\omega_{LO}^2(0) = 2\phi_{11}^{zz}(0) + 4\sum_{N=1}^{\infty}\phi_{11}^{zz}(N)$ and a similar expression for the TO mode. By fitting the series on the right hand sides of (34, 35) to the experimental data of the left-hand sides the range coefficients of the two diagonal elements may be obtained.

To obtain the off-diagonal elements it is necessary to make use of the condition that $Dd$ and $Cd$ are real. This leads to the relations $\text{Im}\{D\} = -\text{Re}\{D\}\tan(\pi\eta)$ and $\text{Im}\{C\} = -\text{Re}\{C\}\tan(\pi\eta)$. Hence:

$$Dd\cos(\pi\eta) = \text{Re}\{D\} \quad \text{and} \quad Cd\cos(\pi\eta) = \text{Re}\{C\} \quad . \quad (36)$$

Further, $D$ and $C$ must vanish at $\eta = 1/2$. From equation (17) this leads to the relation:

$$\phi_{12}^{zz}(0) + \sum_{N=1}^{\infty}\{\phi_{12}^{zz}(N) + \phi_{21}^{zz}(N)\}\cos(\pi N) = 0 \quad , \quad (37)$$

with a similar relation for the xx coefficient and hence:

$$\{\omega_1^2(q) - \omega_2^2(q)\}\cos(\pi\eta) = 4\sum_{N=1}^{\infty}(-1)^N\{\phi_{12}^{zz}(N) + \phi_{21}^{zz}(N)\}\sin^2\{\pi N(1/2 - \eta)\}$$

and

$$\{\omega_5^2(q) - \omega_4^2(q)\}\cos(\pi\eta) = 4\sum_{N=1}^{\infty}(-1)^N\{\phi_{12}^{xx}(N) + \phi_{21}^{xx}(N)\}\sin^2\{\pi N(1/2 - \eta)\} \quad . \quad (38)$$

The first gives $D$, the second $C$. Using the identity $\cos(\pi\eta) = \sin\{\pi(1/2 - \eta)\}$ it is seen that both sides of the above equations vanish at $\eta = 1/2$ as required. A further condition that the range coefficients must obey comes from the $\eta = 0$ value of (38) which leads, noting that $\omega_1(0) = 0$, to

$$\omega_{LO}^2(0) = 4\sum_{N\,\text{odd}}^{\infty}[\phi_{12}^{zz}(N) + \phi_{21}^{zz}(N)] \quad , \quad (39)$$

with a similar expression for the transverse modes. If the Fourier series of the right hand sides of the equations (38) are fitted to the experimental data on the left hand sides multiplied by the factor $\cos(\pi\eta)$, the off-diagonal range coefficients may be obtained. In the two other principal directions of the reciprocal lattice of the hexagonal close packed structure the eigenvectors are functions of the force constants[4] so that for this lattice the method of analysis described above is only able to be used for the [001] direction.

In the diamond lattice[3,22] the eigenfunctions of the modes for $q$ in the [001] direction that are longitudinal at $q = 0$ are of identical form to those of the two longitudinal modes ($i = 1, 2$; LA and LO) in equation (33). The two modes become degenerate at the zone boundary so they may be identified easily and an analysis similar to that used for the hexagonal close packed lattice may be applied to them. In the [110] direction the TO and TA modes depend[3] only on $q$, but with a more complicated dependence, and the same type of analysis may be possible. For the [111] direction[23] although the amplitudes of all the atomic motions are independent of both the force constants and of *q* the phases are not, and it is not clear whether any simplification can be made in this case.

**7. Dynamics and Thermodynamics of Disordered Materials**

Although this paper has so far been concerned with perfectly periodic solids, the main algebraic method that is used in it, the invariance of the trace of a matrix under a change of basis, may be applied to a limited extent to disordered harmonic solids despite the substantial loss of symmetry in these materials. Consider a disordered harmonic lattice, see Weaire and Taylor[24] for a general review, containing *N* non-identical atoms situated at position $R_r$ with mass $M_r$, $r = 1$ to $N$. The equation of motion for the atom at $R_r$ is:

$$M_r \frac{\partial^2}{\partial t^2} u^\alpha(r,t) = -\sum_s \sum_\beta \Phi^{\alpha\beta}(r,s) u^\beta(s,t) \quad , \tag{40}$$

where $u^\alpha(s,t)$ is the displacement of atom *s* in the $\alpha$ direction at time *t* from its equilibrium position. The $\Phi^{\alpha\beta}$ are the harmonic force constants; unlike in a periodic solid these are all different and have no symmetry properties except $\Phi^{\alpha\beta}(r,s) = \Phi^{\beta\alpha}(s,r)$. If the expression

$$u^\alpha(r,t) = \frac{1}{\sqrt{M_r}} u^\alpha(r) e^{-i\omega t} \quad , \tag{41}$$

is substituted into equation (40) the following eigenvalue equation is obtained:

$$\sum_s \sum_\beta \{ \frac{\Phi^{\alpha\beta}(r,s)}{\sqrt{M_r M_s}} - \omega^2 \delta(\alpha,\beta)\delta(r,s) \} u^\beta(s) = 0 \quad . \tag{42}$$

The real eigenvalues of this Hermitian matrix give the squares of the 3*N* vibrational eigenfrequencies $\omega_i^2$, $i = 1$ to $3N$; the eigenvectors $u_i^\alpha(s)$, through equation (41), provide the amplitude and phase of the vibrational motion of atom *r* in mode *i*. From the theorem that the sum of the eigenvalues of a matrix is equal to its trace it follows that:

$$\sum_{i=1}^{3N} \omega_i^2 = \sum_\alpha \sum_r \Phi^{\alpha\alpha}(r,r) / M_r \quad . \tag{43}$$

Using the disordered version of equation (10), we get

$$\Phi^{\alpha\beta}(r,r) = -\sum_s{}' \Phi^{\alpha\beta}(r,s) \quad , \tag{44}$$

where the prime means that the term $s = r$ is omitted from the sum. In this way, we can express the right-hand side of (43) in terms of the more familiar *inter*atomic force constants as

$$-\sum_\alpha \sum_r \sum_s{}' \Phi^{\alpha\alpha}(r,s) / M_r \quad . \tag{45}$$

Consider now the simple model of the lattice regarded as a collection of Einstein oscillators[25] such that each atom is considered to oscillate on its own, the other atoms remaining stationary. This implies that in equation (40) $u^\beta(s,t)$ is non-zero only for $s = r$ so

$$M_r \frac{\partial^2}{\partial t^2} u^\alpha(r,t) = -\sum_\beta \Phi^{\alpha\beta}(r,r) u^\beta(r,t) \quad . \quad (46)$$

The eigenvalue equation for atom $r$ now becomes:

$$\sum_\beta \{\frac{\Phi^{\alpha\beta}(r,r)}{M_r} - \omega^2 \delta(\alpha,\beta)\} u^\beta(r) = 0 \quad . \quad (47)$$

This can be diagonalised to give the squared Einstein eigenfrequencies $(\omega^E_{r,j})^2$, three ($j = 1, 2, 3$) for each different atom $r$; there are $3N$ modes in all. Summing over them all we obtain:

$$\sum_{j=1}^{3} \sum_{r=1}^{N} (\omega^E_{r,j})^2 = \sum_\alpha \sum_r \Phi^{\alpha\alpha}(r,r)/M_r \quad , \quad (48)$$

Noting the equality of the right hand sides of equations (43) and (48) it is seen that

$$\sum_{j=1}^{3} \sum_{r=1}^{N} (\omega^E_{r,j})^2 = \sum_{i=1}^{3N} (\omega_i)^2 \quad , \quad (49)$$

or the sum of the squares of the Einstein frequencies is equal to the sum of the squares of the exact frequencies. This is true for any harmonic lattice, periodic or disordered.

The high temperature expansion for the harmonic lattice specific heat $C(T)$ at temperature $T$ is[1,26]

$$\frac{C(T)}{3Nk} = 1 - \frac{1}{36N}(\frac{\hbar}{kT})^2 \sum_{i=1}^{3N}(\omega_i)^2 + O[T^{-4}] \quad , \quad (50)$$

where the $\omega_i$ are the exact eigenfrequencies. For a periodic solid the $i$ incorporate the wave-vector as well as the branch index, for the disordered solid the $i$ refer to the $3N$ separate vibrational frequencies and for an Einstein solid the $i$ refer to the 3 vibrational frequencies of every atom. Because of relation (47) it can be seen that the Einstein model of lattice dynamics gives exactly not only the constant high temperature limit of the specific heat, $3Nk$, but also the next term (in $T^{-2}$) in its high temperature expansion. Accordingly, the Einstein model gives exactly to order $T^{-3}$ the high temperature expansion for the specific heat[27-29]. This result may be useful in the numerical computation of the high temperature thermodynamic properties of disordered lattices because it is quicker to diagonalise $N^2$ 3x3 matrices than one $3N$x$3N$ matrix[30].

## 8. Conclusion

A general derivation of the of the Forman-Lomer theorem has been presented and it has been shown that from simple combinations of certain phonon frequencies it is possible to obtain detailed information about the range of interatomic forces in crystals. The results apply to some non-primitive as well as primitive lattices. A general statement of the Blackman Sum Rule has been obtained and allows a quantitative analysis, in terms of the atomic force constants, of deviations from this rule. It is also shown that the sum of the squares of the exact frequencies is equal to the sum of the squares of the Einstein frequencies, a result which may be useful for the computation of the high temperature properties of disordered harmonic lattices.